\documentclass{jfm}
\usepackage{graphicx}
\usepackage{epstopdf,epsfig}
\usepackage{color}
\usepackage{amsmath}
\usepackage{soul}
\newcommand{\ct}[1]{\multicolumn{1}{c}{#1}}

\shorttitle{Near-wall turbulence modulation by small inertial particles}
\shortauthor{P. Costa, L. Brandt and F. Picano}

\title{Near-wall turbulence modulation by small inertial particles}

\author{Pedro Costa\aff{1}
\corresp{\email{p.simoes.costa@gmail.com}},
Luca Brandt\aff{2}
\and Francesco Picano\aff{3}}

\affiliation{
\aff{1}Faculty of Industrial Engineering, Mechanical Engineering and Computer Science, University of Iceland, Hjardarhagi 2-6, 107 Reykjavik, Iceland\\
\aff{2}Linn\'e FLOW Centre and SeRC (Swedish e-Science Research Centre), Department of Engineering Mechanics, KTH, SE-100 44 Stockholm, Sweden\\
\aff{3}Department of Industrial Engineering, University of Padova, Via Venezia, 1, 35131, Padova, Italy}

\begin{document}

\maketitle

\begin{abstract}
We use interface-resolved simulations to study near-wall turbulence modulation by small inertial particles, much denser than the fluid, in dilute/semi-dilute conditions. We considered three bulk solid mass fractions, $\Psi=0.34\%$, $3.37\%$ and $33.7\%$, with only the latter two showing turbulence modulation. The increase of the drag is strong at $\Psi=3.37\%$, but mild in the densest case. Two distinct regimes of turbulence modulation emerge: for smaller mass fractions, the turbulence statistics are weakly affected and the near-wall particle accumulation increases the drag so the flow appears as a single-phase flow at slightly higher Reynolds number. Conversely, at higher mass fractions, the particles modulate the turbulent dynamics over the entire flow, and the interphase coupling becomes more complex. In this case, fluid Reynolds stresses are attenuated, but the inertial particle dynamics near the wall increases the drag via correlated velocity fluctuations, leading to an overall drag increase. Hence, we conclude that, although particles at high mass fractions reduce the fluid turbulent drag, the solid phase inertial dynamics still increases the overall drag. However, inspection of the streamwise momentum budget in the two-way coupling limit of vanishing volume fraction, but finite mass fraction, indicates that this trend could reverse at even higher particle load.
\end{abstract}

\section{Introduction}\label{sec:intro}
The turbulent channel flow laden with small, gravity-free inertial spheres is a paradigmatic multiphase flow, yet not fully understood and correctly modelled. 
This system has been widely studied using direct numerical simulations (DNS) of the Navier-Stokes equations using the point-particle approximation \citep{Balachandar-and-Eaton-ARFM-2010}, which assumes that interphase coupling is localized at a single point and the particle dynamics driven by an undisturbed flow field, evaluated at the particle position. This approximation allows us to study the flow dynamics beyond the particle scale, without solving the actual flow around each particle. Despite the many studies, the validity of these  particle--fluid coupling models has not been fully assessed for canonical turbulent wall flows, in particular when the underlying turbulence is altered by the dispersed phase.

When the particle feedback on the flow becomes important -- the two-way coupling regime -- care should be taken in the estimation of the undisturbed velocity. The challenge is conciliating the estimation of an undisturbed velocity sampled by the particle with the need of forcing the local velocity field. Indeed, approaches for accurate and realistic two-way coupling methods have been pursued since the work of \cite{Crowe-et-al-JFE-1977}, and are still object of active research \cite[see, e.g.,][]{Gualtieri-et-al-JFM-2015,Horwitz-and-Mani-JCP-2016,Ireland-and-Desjardins-JCP-2017,Battista-et-al-JFM-2019,Pakseresht-and-Apte-JCP-2020}. Moreover, results in the literature for integral observables show qualitatively different trends. For instance, while \cite{Vreman-JFM-2007,Zhao-et-al-PoF-2010} have observed a drag-reducing behaviour in two-way coupled particle-laden turbulent channel flow, other studies have measured a drag increase \citep{Battista-et-al-JFM-2019}. These opposite trends suggest the presence of different regimes of momentum transfer in wall-bounded particle-laden turbulence. Indeed, \cite{Capecelatro-et-al-JFM-2018} used Euler--Lagrange volume-filtered DNS to investigate in detail the turbulent kinetic energy (TKE) budget in vertical particle-laden channel flow at different volume fractions. Different regimes were observed with increasing mass fraction -- first, a turbulent regime at low mass fractions; second, increasing the mass fraction, the turbulence activity decreases until the flow laminarizes, due to the growing importance of a ``drag dissipation-and-exchange term'' associated with  fluid--particle correlated velocity fluctuations; finally, increasing the mass fraction beyond this limit triggers a second mechanism for TKE increase, due to the average velocity difference between phases, which grows important and re-energizes fluid turbulence. Despite the exciting developments exploiting particle-modelled two-way coupling interactions, there is a clear need for high-fidelity data to support the body of work using these modelled simulations, and to possibly reconcile seemingly conflicting observations of drag increase and drag reduction in two-way coupling point-particle numerical simulations.\par
Indeed, the tremendous developments of approaches for DNS with modelled particle--fluid coupling have not been accompanied by the high-fidelity data which are essential for validating the underlying assumptions. These data can be obtained either from interface-resolved simulations of the Navier-Stokes equations, or well-controlled experiments matching the computational set-up. Nowadays, the first direct comparisons between point-particle DNS data and experiments or interface-resolved DNS are possible. Examples are the work by \cite{Wang-et-al-IJMF-2019}, which compares one-to-one two-way coupled point-particle DNS with experimental measurements for vertical turbulent channel flow at different volume loads, and the interface-resolved (also denoted particle-resolved) DNS of spherical particles near the point-particle limit in homogeneous isotropic turbulence \citep{Schneiders-et-al-JFM-2017,Mehrabadi-et-al-JFM-2018,Frohlich-et-al-FTC-2018} and in turbulent channel flow \citep{Horne-and-Mahesh-JCP-2019,Costa-et-al-JFM-2020}.\par
The present work also exploits interface-resolved simulations and focuses on near-wall  turbulence modulation by small inertial particles. 
We adopt the configuration in \cite{Costa-et-al-JFM-2020} and add results at higher volume fraction. Three bulk mass fractions $\Psi=0.34\%$, $3.37\%$ and $33.7\%$ at fixed particle to fluid density ratio ($\Pi_\rho=100$) are hence considered, where the latter two show non-negligible two-way coupling effects. 
Our results reveal two distinct mechanisms for turbulence modulation. At lower volume fractions, the flow is only dense very close to the wall, and the higher flow inertia in this region, with particles travelling much faster than the fluid, results in a drag-increased flow resembling single-phase turbulence at slightly higher Reynolds number. At higher volume fractions, the dispersed phase is dynamically important over the entire channel. Here, the drag increasing effect is amplified by correlated particle velocity fluctuations, but counterbalanced by a substantial fluid turbulence drag reduction. 
This results in a milder increase in drag for the denser case. These observations may be explained in light of the streamwise momentum balance for vanishing volume fraction, but with non-negligible mass fraction.
 
\section{Methods and Computational Set-up}\label{sec:methods}

Since we use here the tools and set-up in \cite{Costa-et-al-JFM-2020}, with one additional case at the largest volume fraction, we briefly summarize the numerical method and refer to the previous work for more details. We solve the continuity and Navier-Stokes equations for an incompressible Newtonian fluid, together with the Newton-Euler equations driving the motion of the solid spherical particles. These two sets of equations are coupled directly using the immersed-boundary method developed by \cite{Breugem-JCP-2012} \cite[see also][]{Uhlmann-JCP-2005}, built on a standard second-order finite-difference method on a three-dimensional, staggered Cartesian grid, using a fast-Fourier-transform-based pressure-projection method \citep{Kim-and-Moin-JCP-1985,Costa-CAMWA-2018}. Short-range particle--particle/particle--wall interactions (lubrication and solid--solid contact) are modelled using the method of \cite{Costa-et-al-PRE-2015}, as in \cite{Costa-et-al-JFM-2020}. More specifically, as regards solid--solid contact, the particles are frictionless, and with a normal solid--solid coefficient of restitution of $0.97$.\par
Turbulent channel flow is simulated in a domain periodic in the streamwise ($x$) and spanwise ($z$) directions, with no-slip and no-penetration boundary conditions imposed at the walls ($y=h\mp h$), where $h$ is the channel half-height. The flow is driven by a uniform pressure gradient that ensures a constant bulk velocity. The bulk Reynolds number is equal to $\Rey_b = U_b(2h)/\nu=5\,600$, which corresponds to an unladen friction Reynolds number $\Rey_\tau^{sph} = u_\tau h/\nu\approx 180$; where $U_b$ is the bulk flow velocity and $u_\tau$ the wall friction velocity. The particle properties are chosen to yield a particle Reynolds number $\Rey_p = D u_\tau/\nu =D^+= 3$, and Stokes number $St_p = \Pi_\rho \Rey_p^{2}/18=50$, where $\nu$ is the fluid kinematic viscosity, $D$ is the particle diameter and $\Pi_\rho$ the particle-to-fluid mass density ratio; since the particle size is restricted by resolution requirements, $\Pi_\rho$ is used to achieve the target particle Stokes number, which was demonstrated to feature highly inhomogeneous particle distributions in wall turbulence \cite[see, e.g.,][]{Sardina-et-al-JFM-2012}. Three values of solid volume fraction are considered, increasing by factors of $10$: $\Phi\simeq3\cdot10^{-5}$, denoted very dilute (\textit{VD}), $\Phi\simeq 3\cdot10^{-4}$, dilute (\textit{D}), and $\Phi\simeq 3\cdot10^{-3}$, semi-dilute (\textit{SD}), with the data pertaining to the first two cases also used in the recent study by \cite{Costa-et-al-JFM-2020}. Table~\ref{tbl:comput_params} shows all relevant physical and computational parameters.\par
\begin{table}
  \begin{center}
  \def~{\hphantom{0}}
  \begin{tabular}{clll}
      \ct{Case}   & \ct{$\Phi \,\,\,\,(N_p)$} & \ct{$\Psi$} & Notes  \\[3pt]
      \textit{VD} & $0.003\%\,\,(  500)~~~$   & $0.337\%$   & interface-resolved  (very dilute) \\
      \textit{D~} & $0.034\%\,\,(5\,000)~~$   & $3.367\%$   & interface-resolved  (dilute)      \\
      \textit{SD} & $0.337\%\,\,(50\,000)~$   & $33.67\%$   & interface-resolved  (semi dilute)
  \end{tabular}
      \caption{Computational parameters. $\Phi$/$\Psi$ denote the bulk solid volume/mass fraction, and $N_p$ the total number of particles. Common  to  all cases: bulk Reynolds number $\Rey_b = 5\,600$ (i.e. friction Reynolds number in the single-phase limit $\Rey_\tau^{sph}\approx 180$); particle size ratio $D/(2h)=1/120$; particle-to-fluid mass density ratio $\Pi_\rho=100$. 
  These correspond to a particle Reynolds number based on the unladen reference values of $D^+=3$ and Stokes number $St=50$. The fluid domain is discretized on a regular Cartesian grid with $(L_x/N_x)\times (L_y/N_y)\times (L_z/N_z) = (6h/4320)\times (2h/1440)\times (3h/2160)$, while the particles are resolved with $D/\Delta x=12$ grid points over the particle diameter (in total $420$ Lagrangian grid points) \cite[same as][]{Costa-et-al-JFM-2020}.}
  \label{tbl:comput_params}
  \end{center}
\end{table}
Finally, unless otherwise stated, the mesoscale-averaged profiles reported in this manuscript correspond to intrinsic averages (i.e., averaged only over the corresponding phase) in time and along the two homogeneous directions for each phase \citep{Costa-et-al-JFM-2020}, obtained with wall-parallel bins with thickness equal to the grid spacing.
\section{Results}\label{sec:results}
\begin{figure}
   \centering
   \includegraphics[width=0.49\textwidth]{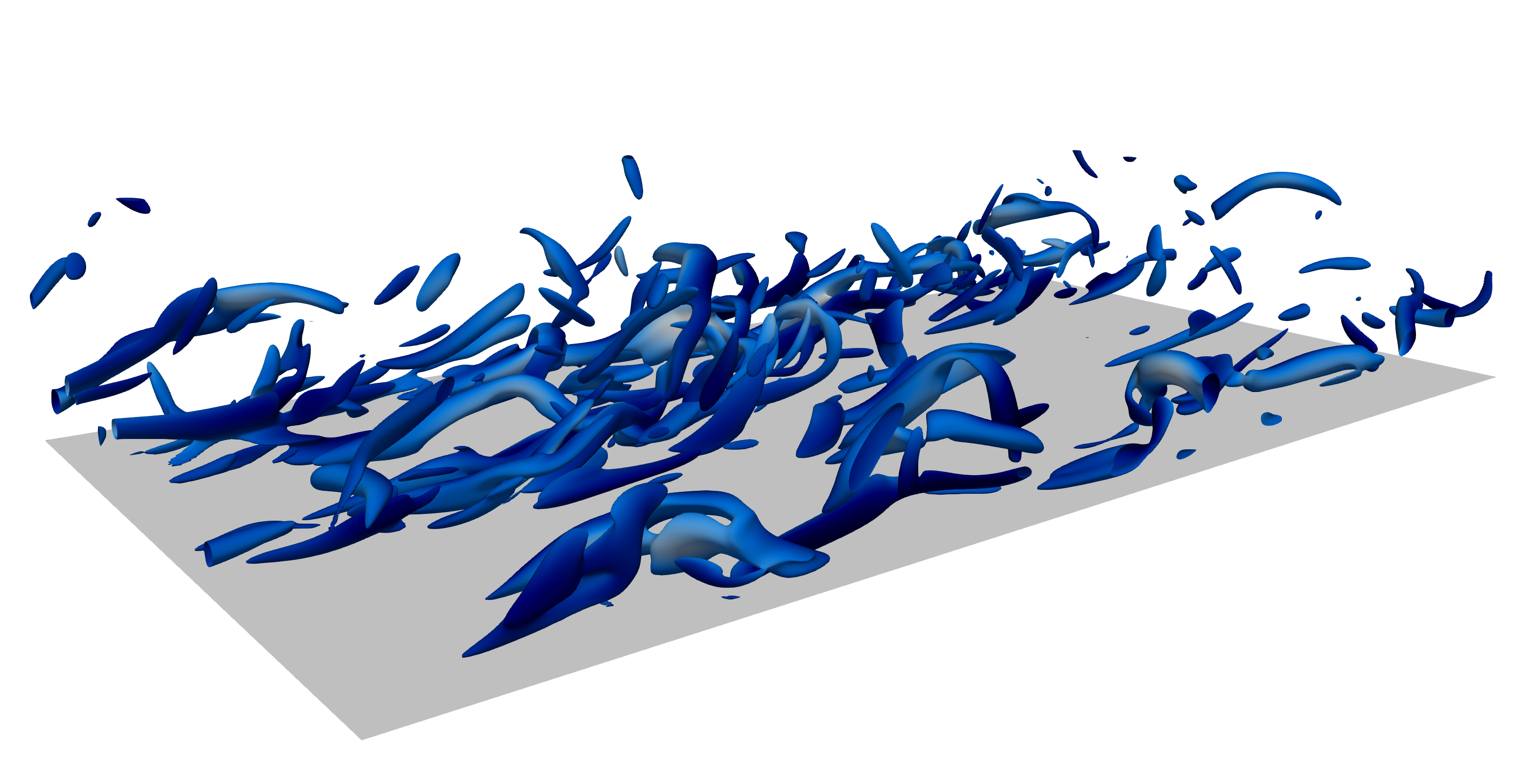}\hfill
   \includegraphics[width=0.49\textwidth]{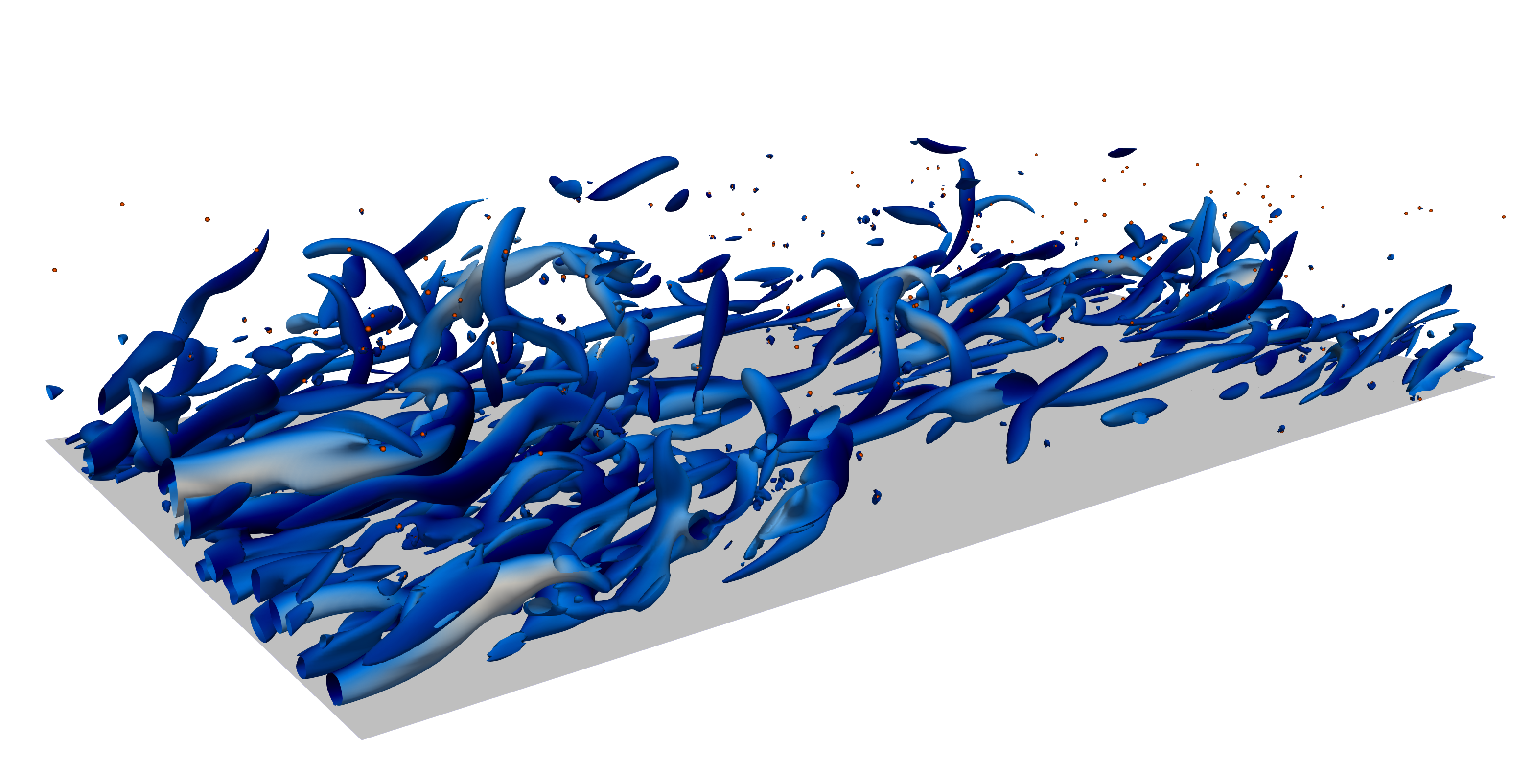}
   \put(-270, 15){single-phase}
   \put( -70, 15){case \textit{VD}}\\
   \includegraphics[width=0.49\textwidth]{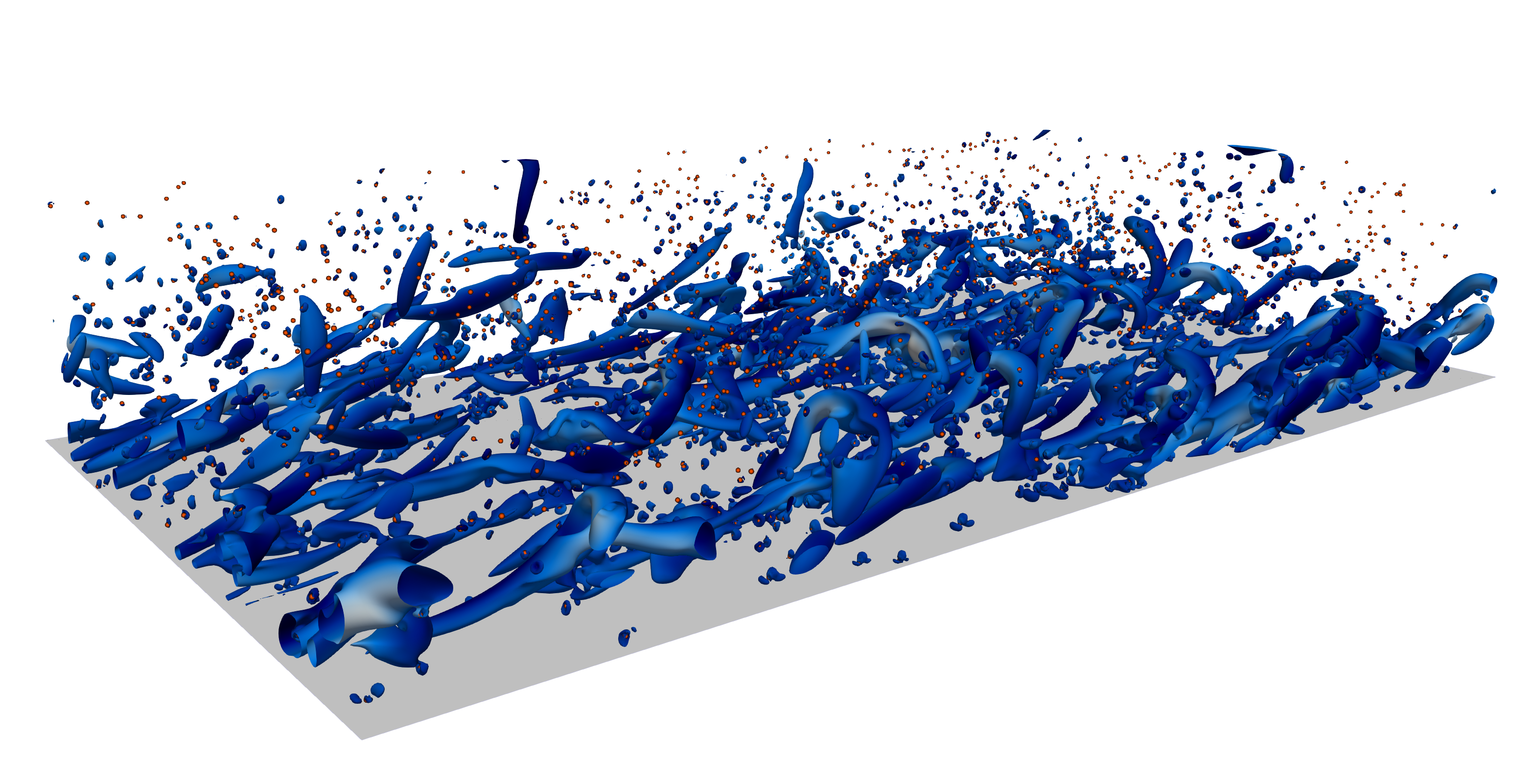}\hfill
   \includegraphics[width=0.49\textwidth]{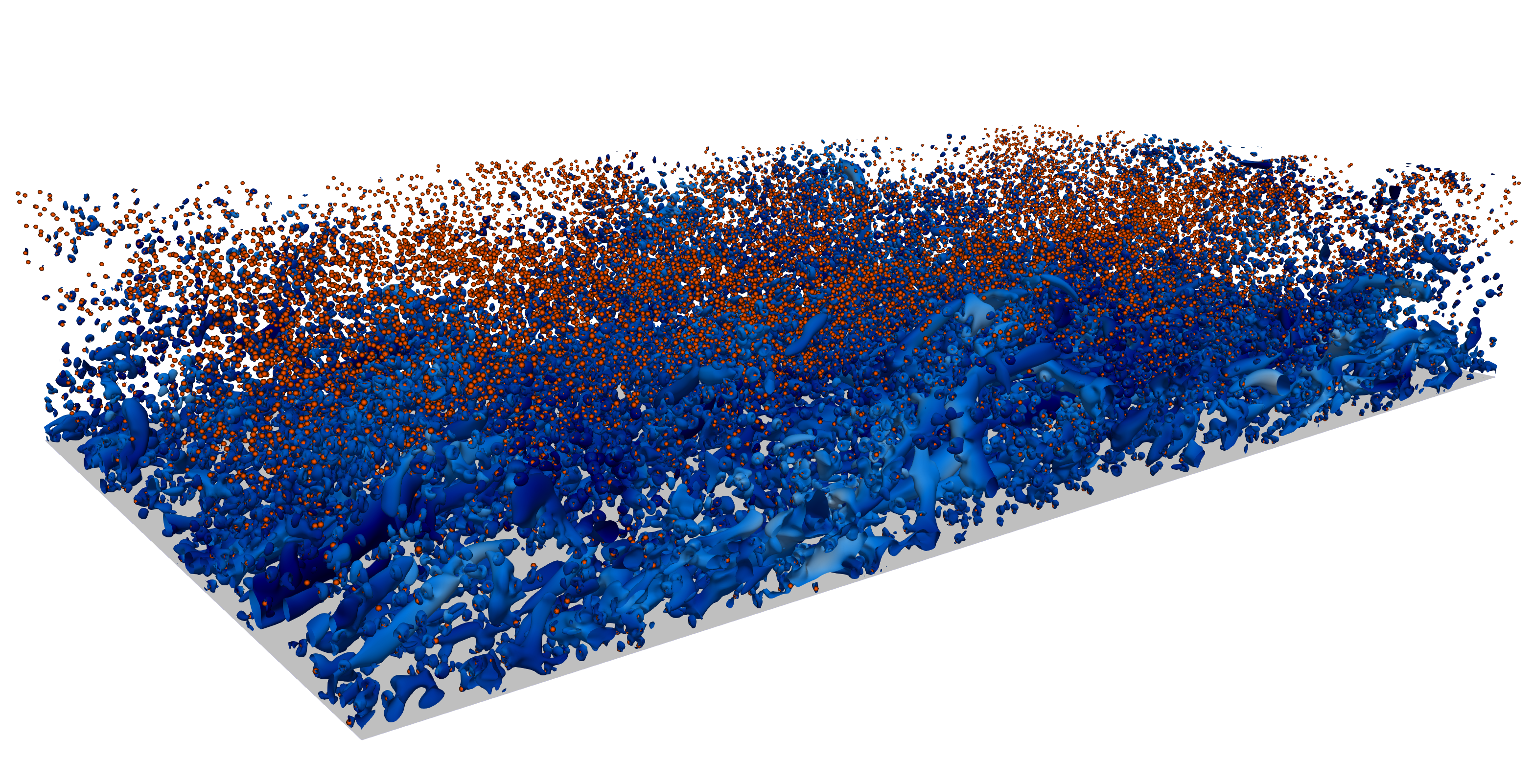}
   \put(-270, 15){case \textit{ D}}
   \put( -70, 15){case \textit{SD}}
    \caption{Three-dimensional flow visualizations of the different cases under consideration. We display surfaces of constant second invariant of the velocity gradient tensor $Q=20(U_b/h)^2$, coloured by the local wall-normal velocity (white -- high and blue -- low). The particles are shown to scale, in orange colour. See table~\ref{tbl:comput_params} for a description of the different cases.}\label{fig:visus_3d}
\end{figure}

A three-dimensional visualization of the flow pertaining to the different cases is reported in figure~\ref{fig:visus_3d}, where iso-surfaces of positive second invariant of the velocity gradient tensor $Q$ (coloured by the local wall-normal velocity) and the particles are displayed. High-vorticity spots, footprints of the presence of the particles, are noticeable in all particle-laden cases, especially closer to the wall, where, as we will see, the mean slip velocity between the phases is highest. As expected, no significant qualitative differences can be seen between the single-phase and the very dilute case \textit{VD}, apart from the very small number of dispersed particles in the latter case. 
Qualitative differences between these cases and case \textit{D} are also small; however, an increased number of high-vorticity spots due to the particles is obvious over the entire domain. Finally, a strong modulation of the flow dynamics by the particles is seen when inspecting case \textit{SD}, where the disruption of the flow coherent structures is evident. One of the main messages of the present work is that, although cases \textit{D} and \textit{SD} have macroscopic flow variations and, therefore, are formally in a two-way coupling regime, the mechanism of turbulence modulation is definitively different.
\par
\begin{figure}
   \centering
   \includegraphics[width=0.49\textwidth]{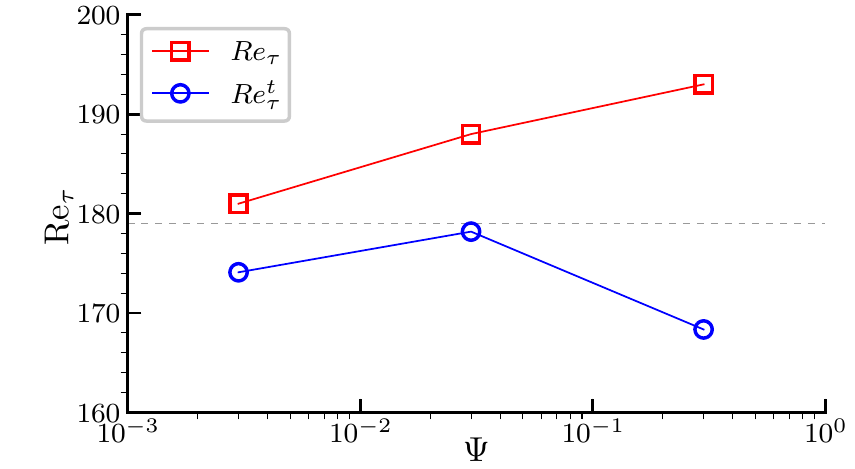}\hfill
   \includegraphics[width=0.49\textwidth]{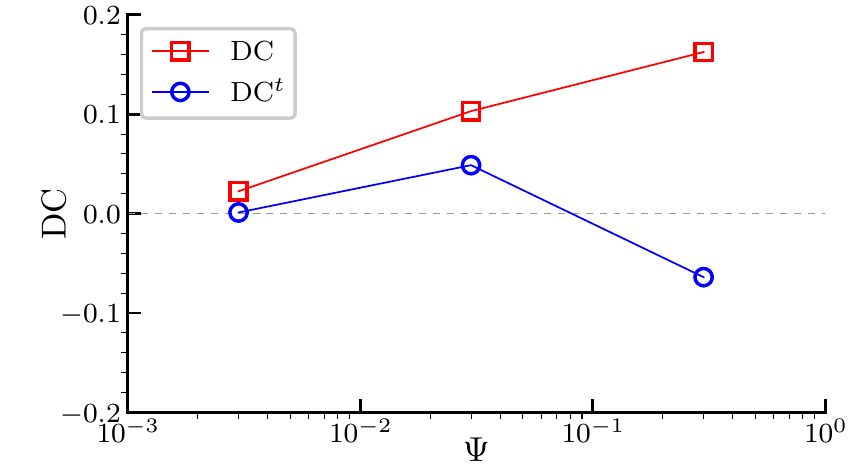}
    \put(-380, 105){\small(\textit{a})}
    \put(-187, 105){\small(\textit{b})}
    \caption{(\textit{a}) Friction Reynolds number $\Rey_\tau$  versus the bulk volume fraction $\Psi$; $\Rey_\tau^t$ is based on a velocity scale computed from the slope of the outer-scaled Reynolds shear stresses evaluated at the channel centreline $u_\tau^{t} = \sqrt{\partial_{y/h} \left\langle u^\prime v^\prime \right\rangle |_{y/h=1}}$, as a rough measure of the contribution of fluid velocity correlations to the mean wall shear \cite[see][]{Picano-et-al-JFM-2015}. (\textit{b}) Drag modulation $\mathrm{DC}=\Rey_\tau^2/\Rey_\tau^{2,sph}-1$, computed with $\Rey_\tau$ and $\Rey_\tau^t$ ($\mathrm{DC}^t$).}\label{fig:retau}
\end{figure}

Figure~\ref{fig:retau}(\textit{a}) presents the friction Reynolds number as a function of the bulk volume fraction, while the change in drag relative to the unladen flow is quantified in panel (\textit{b}). The friction Reynolds number increases significantly ($5\%$) from case \textit{VD} to case \textit{D}, despite the relatively low mass fraction $\Psi = \mathcal{O}(10^{-2})$. As discussed in \cite{Costa-et-al-JFM-2020}, this increase is attributed to the higher inertia induced by the large local mass fraction near the wall; the particles are driven by turbophoresis towards the wall,  where they experience a large apparent slip velocity. Further increasing the mass fraction by an order of magnitude (case \textit{SD}) results in a  mild increase in drag, only about $3\%$ with respect to case \textit{D}. This milder increase indicates that a competing drag-reducing mechanism comes into play at higher mass loading. The blue symbols in figure~\ref{fig:retau} display quantities computed using a wall friction velocity determined from the centreline slope of the Reynolds stress, so as to quantify the decrease in turbulent momentum transfer, as discussed in detail later.\par
\begin{figure}
   \centering
   \includegraphics[width=0.49\textwidth]{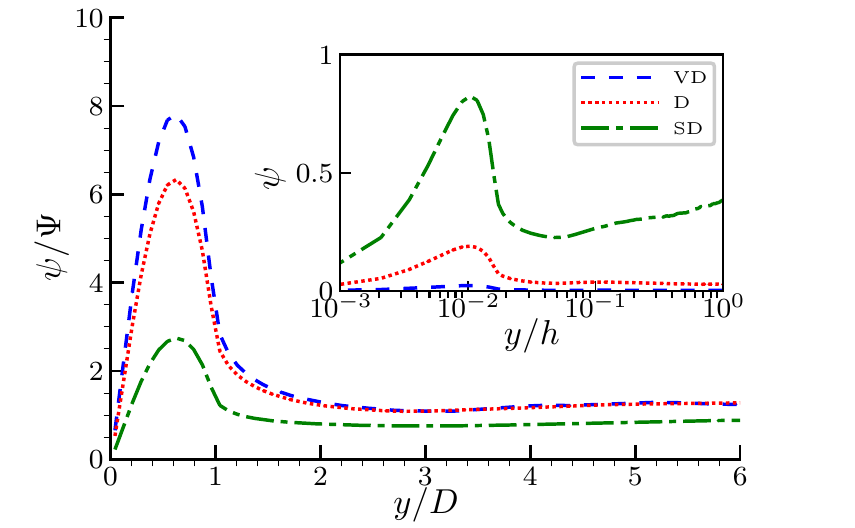}\hfill
   \includegraphics[width=0.49\textwidth]{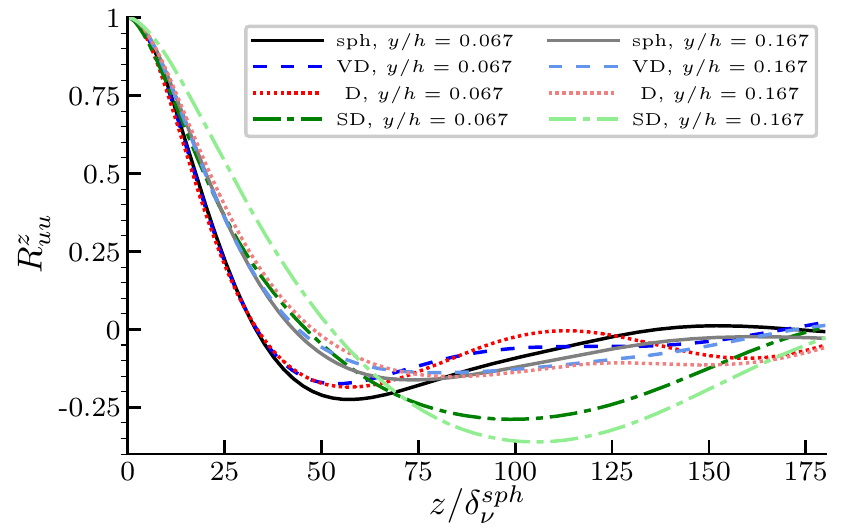}
    \put(-380, 105){\small(\textit{a})}
    \put(-187, 105){\small(\textit{b})}
    \caption{(\textit{a}) Local solid mass fraction $\psi$, normalized by the bulk value $\Psi$ as a function of the wall-normal distance in particle diameters (note that the profile is clamped beyond $y/D=6$, to highlight the near-wall peaks); the inset shows the nominal (i.e., not normalized) local mass fraction $\psi$ as a function of the outer-scaled wall-normal distance. The non-zero slope at the centreline, for case \textit{SD}, is attributed to insufficient statistical convergence in this region where the flow is more dilute and the volume fraction fluctuations higher, further highlighted by the logarithmic scaling. (\textit{b}) Outer-scaled spanwise autocorrelation of streamwise velocity at $y/h=0.067$  and $y/h=0.167$, ($y/\delta_{\nu}^{sph}\approx 12$ and $30$); `sph' corresponds to the unladen flow.}\label{fig:phase}
\end{figure}

Figure~\ref{fig:phase}(a) shows the mass fraction normalized by the corresponding bulk value versus the wall-normal distance in particle diameters. All the cases show a peak  at the wall, corresponding to a particle layer. Note also that these peaks occur slightly beyond a wall distance of one particle radius, due to the bouncing dynamics caused by particle--wall collisions. Clearly, the fraction of particles near the wall decreases with increasing mass loading. This decrease in wall accumulation is attributed to the increased shear rate near the boundary, which enhances lift forces \citep{Costa-et-al-JFM-2020}.
However, case \textit{SD} shows a disproportionally strong decrease of the relative particle concentration, considering the relatively low increase in wall shear with respect to case \textit{D}. This is partly caused by two-way coupling effects, which dampen the intensity of the wall-normal fluid velocity fluctuations responsible -- to first approximation -- for the turbophoretic drift \citep{Marchioli-and-Soldati-JFM-2003}. In addition, particle--particle interactions may become significant. The inset of figure~\ref{fig:phase}(\textit{a}) shows the mass fraction profile as a function of the outer-scaled wall-normal distance. While the mass fraction profiles become uniform far from the wall for cases \textit{VD} and \textit{D}, the distribution  shows a mild monotonic increase with the wall distance in  case \textit{SD}. These observations indicate that, for this densest case, non-negligible particle--particle interactions may be driving particles towards regions of low shear. The mechanism underlying this drift is most likely similar to that reported in \cite{Fornari-et-al-PoF-2016} -- particle inertia and local high shear promote inter-particle interactions, causing a net  migration towards low shear regions. We attribute the non-zero slope at the channel centreline of the profile pertaining to case \textit{SD}  to a lower statistical convergence in the channel core in combination with the amplification induced by the logarithmic scale, as explained in the figure caption.
\par

\begin{figure}
   \centering
   \includegraphics[width=0.89\textwidth]{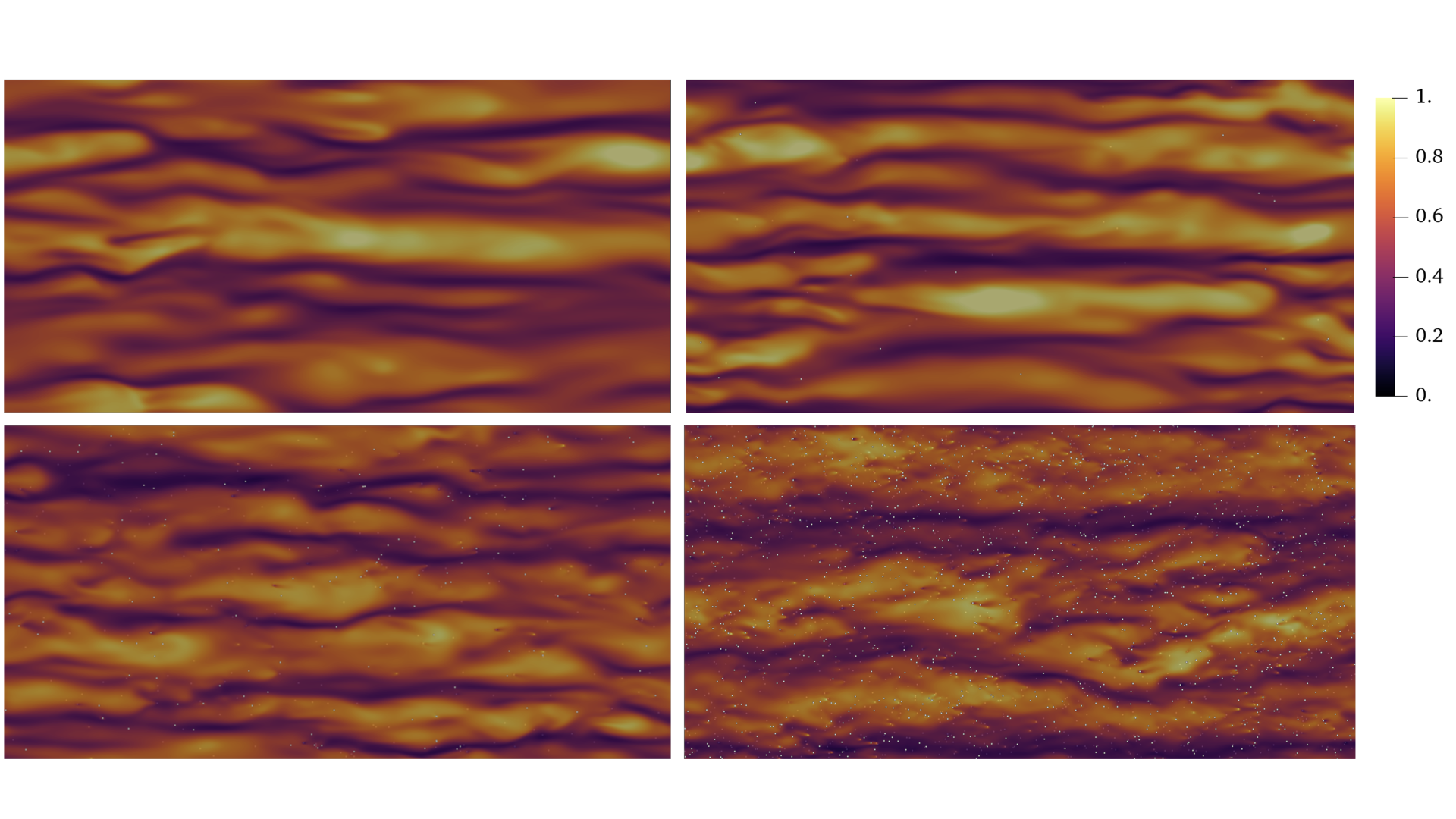}
   \put(-225,100){\color{white}\tiny single-phase}
   \put( -50,100){\color{white}\tiny case \textit{VD}}
   \put(-208, 20){\color{white}\tiny case \textit{D}}
   \put( -50, 20){\color{white}\tiny case \textit{SD}}
   \put(-20,175){\tiny $u$}
   \caption{Contours of outer-scaled streamwise fluid velocity $u$ at a plane $y/h = 0.07$, together with the particles located at $y/h<0.15$. Perspective of an observer aligned with the negative wall-normal direction, with flow from left to right (the contour is shown with slight transparency).}\label{fig:visu_planes}
\end{figure}
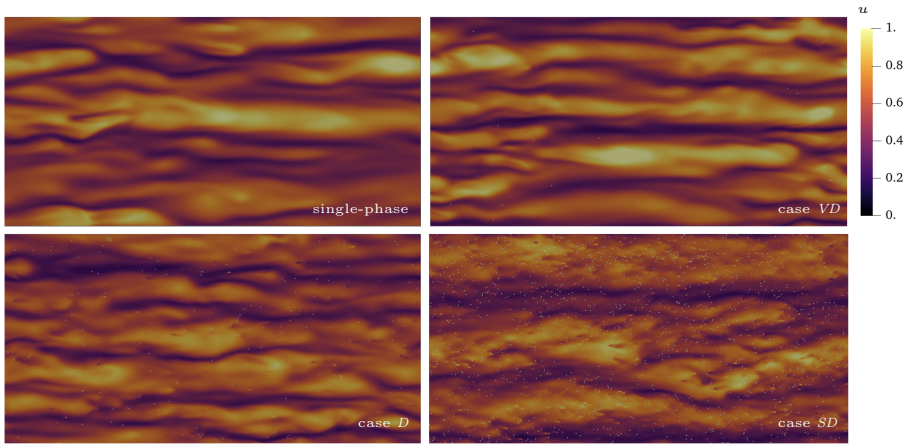

The inset of figure~\ref{fig:phase}(\textit{a}) also illustrates where two-way coupling effects are expected to be important. These can be envisaged near the wall for the less dilute cases, as the solid mass fraction increases to $20\%$ for case \textit{D}, and to about $80\%$ for case \textit{SD}. Away from the wall, instead,  the solid mass fraction  retains high values only for case \textit{SD},  $25\%-40\%$, and therefore turbulence modulation is also expected in the bulk, which we will relate to the milder overall drag increase from case \textit{D} to case \textit{SD}. In the latter case, we observe a significant turbulence modulation by the solid particles: the spacing between low- and high-speed streaks is larger, the spanwise velocity variation smoother, and the maximum streak amplitude occurs at a much larger wall-normal distance, which is reflected in the spanwise autocorrelation of the streamwise velocity in wall-parallel planes shown in figure~\ref{fig:phase}(\textit{b}). This trend can be clearly seen in figure~\ref{fig:visu_planes}, showing the contours of the near-wall streamwise fluid velocity.
Case \textit{SD} also presents non-negligible four-way coupling effects. In particular, we estimated the particle collision frequency near the wall following \cite{Sundaram-and-Collins-JFM-1997}:
\begin{equation}
    N_c = \frac{1}{2}\pi R^2n^2 g(R)\left\langle \Delta U^{n-}\right\rangle \mathrm{,}\label{eqn:col_estim}
\end{equation}
where $R$ is the particle radius, $n$ the local number density, $g(R)$ the particle radial distribution function at the imminence of contact and $\left\langle \Delta U^{n-}\right\rangle $ the mean particle relative velocity projected along the line-of-centres, conditioned to negative values (i.e.\ promoting a collision) at the imminence of contact (i.e.\ interparticle distance equal to one particle diameter).
We found the collision frequency for case \textit{SD} to be $5.5$ near the wall and to reduce to $0.53$ in the bulk, a quantity defined per unit volume $L_x L_z D$ and bulk eddy turnover time $(O(h/u_\tau))$, more than one order of magnitude larger than in case \textit{D} ($0.29$ near the wall and $0.002$ in the bulk). We should note that, owing to the computational cost and the low volume fractions considered, we were unable to collect enough data to directly measure particle collision statistics. Also, a direct assessment of collision frequency can yield deceiving results due to the possibility of sustained contacts; see \cite{Kuerten-and-Vreman-IJMF-2016}.\par
\begin{figure}
   \centering
   \includegraphics[width=0.49\textwidth]{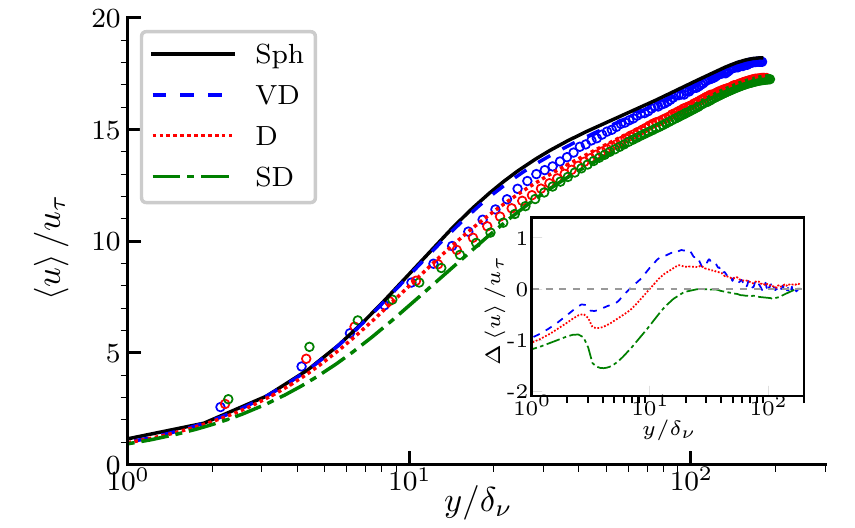}\hfill
   \includegraphics[width=0.49\textwidth]{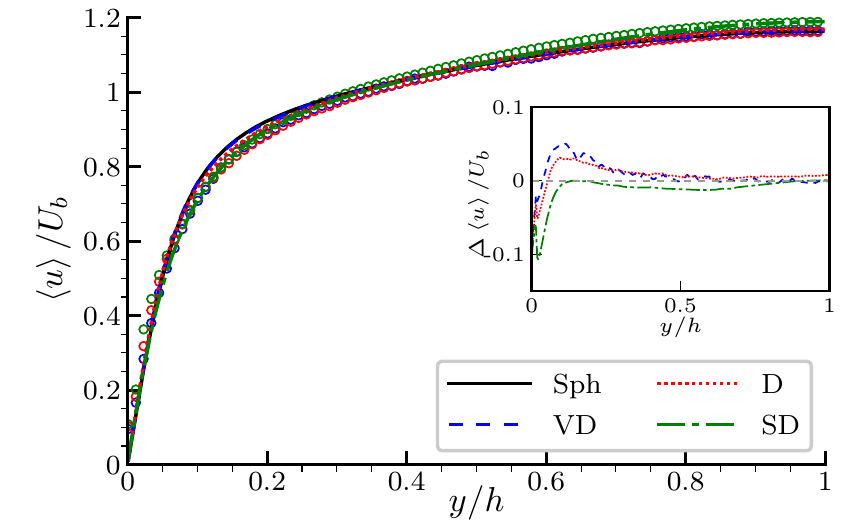}
    \put(-380, 105){\small(\textit{a})}
    \put(-187, 105){\small(\textit{b})}
    \caption{Inner- (\textit{a}) and outer-scaled (\textit{b}) mean streamwise velocity profiles. Lines -- fluid velocity; symbols -- particle velocity. The insets show the apparent fluid-to-particle slip velocity, defined as the difference between the intrinsic mean velocity of each phase.}\label{fig:umean}
\end{figure}

The inner-scaled mean velocity profiles, see figure~\ref{fig:umean}(a), resemble those of the single-phase flow, with a downward shift that indicates drag increase (the difference is less apparent in the corresponding outer-scaled profiles shown in panel (b)). 
The inset of the figure shows the particle-to-fluid apparent slip velocity, defined as the difference between the mean velocity profiles of the fluid and solid phase. 
The negative minimum in the near-wall region is due to the higher particle velocity  where the fluid velocity is vanishing. The local maximum of positive slip velocity occurs in the buffer region, which is an indirect indication of the well-known tendency of near-wall inertial particles to oversample regions of low streamwise fluid velocity observed in numerous studies \cite[e.g.,][among others]{Rouson-and-Eaton-JFM-2001,Marchioli-and-Soldati-JFM-2003}. The higher the mass loading, the weaker this slip velocity is. Case \textit{SD}, in particular, shows a deeper minimum of the slip velocity, and virtually no slip between the two phases beyond $y>10\delta_\nu$. This deeper minimum is reflected in more pronounced wakes for particles moving faster than the fluid, see figure~\ref{fig:visu_planes}. Interestingly, the fluid-to-particle apparent slip velocity does not attain positive values for case \textit{SD}, meaning that particles are less prone to sampling low-speed regions.
Since the low (and high) speed regions are located further away from the wall, particles departing from the wall may not reside for long enough in these regions. Finally, we should note that wider streak spacing with larger wall-normal extent is often a feature of turbulent drag reduction \cite[see e.g.][]{Tiederman-et-al-JFM-1985}.\par
\begin{figure}
   \centering
   \includegraphics[width=0.49\textwidth]{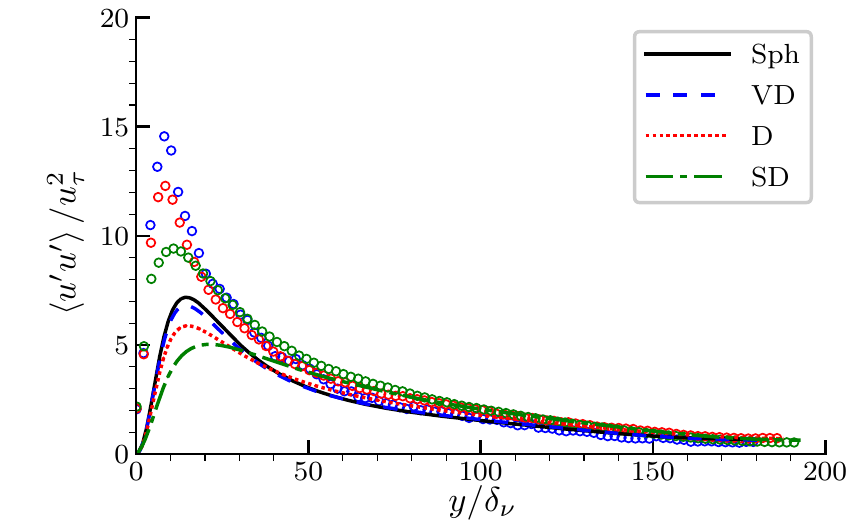}\hfill
   \includegraphics[width=0.49\textwidth]{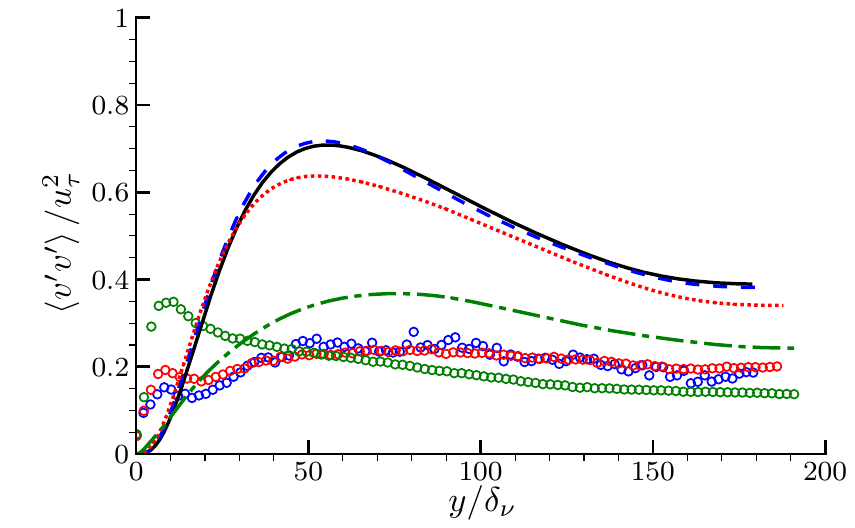}
    \put(-380, 105){\small(\textit{a})}
    \put(-187, 105){\small(\textit{b})}\\
   \includegraphics[width=0.49\textwidth]{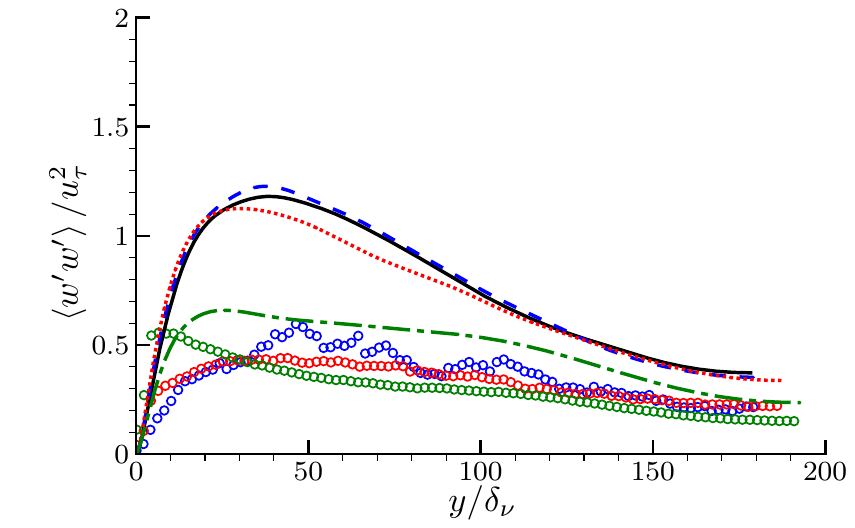}\hfill
   \includegraphics[width=0.49\textwidth]{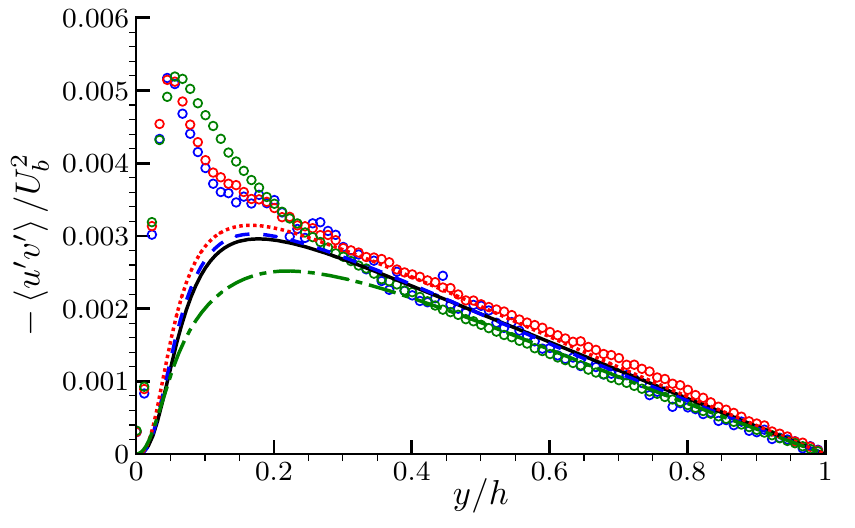}
    \put(-380, 105){\small(\textit{c})}
    \put(-187, 105){\small(\textit{d})}
    \caption{Second-order moments of mean fluid (lines) and particle (symbols) velocity. (\textit{a}) Inner-scaled streamwise, (\textit{b}) wall-normal and (\textit{c}) spanwise velocity variances, and (\textit{d}) outer-scaled Reynolds stresses profile. The legend in (\textit{a}) holds for all panels.}\label{fig:velstats}
\end{figure}\par
Figure~\ref{fig:velstats} presents the second-order moments of the fluid and particle velocities. Particle fluctuations are usually higher in the near-wall region and smaller or similar to that of the fluid in the bulk. The modifications with respect to the unladen case are much smaller for cases \textit{VD} and \textit{SD}, compared with case \textit{SD}. As discussed in \cite{Costa-et-al-JFM-2020}, the minor deviations of \textit{VD} with respect to the unladen case are attributed to a lower statistical convergence of the massive particle-resolved simulation, and a slight two-way coupling effect, while the more pronounced differences in case \textit{D} are caused by the increase in mean wall shear. Compared with these two cases, case \textit{SD} shows three significant differences: a strong reduction of the streamwise velocity fluctuations near the wall and enhancement in the bulk, a large decrease of the wall-normal and spanwise velocity variances across the channel, and a reduction in Reynolds shear stresses. This observation is consistent with particle-modelled DNS and experiments of particle-laden wall-bounded turbulence \cite[see, e.g.][]{Kulick-et-al-JFM-1994,Zhao-et-al-PoF-2010,Richter-PoF-2015,Capecelatro-et-al-JFM-2018,Wang-et-al-IJMF-2019}. Focusing on the fluid Reynolds shear stress, case \textit{D} features a slight increase of the peak value and slope in the outer region, while the opposite applies to case \textit{SD}. Note that the particle Reynolds stresses are always larger than the fluid ones.\par
Despite the overall increase in drag, the significant decrease of the Reynolds shear stresses for case \textit{SD} is consistent with the reduced contribution to the overall drag from the fluid turbulence  shown in figure~\ref{fig:retau} by the friction Reynolds number based on the velocity based on the slope of the Reynolds stresses, $u_\tau^{t} = \sqrt{\partial_{y/h} \left\langle u^\prime v^\prime \right\rangle |_{y/h=1}}$. This quantity initially increases with the mass loading (case \textit{VD} to \textit{D}), denoting an enhancement of turbulence and its induced drag. Further increasing the solid mass fraction, (case \textit{SD}),  $Re_{\tau}^t$ decreases -- even below the single-phase flow --  while the conventional friction Reynolds number increases. In other words, while turbulence is attenuated, a fundamentally different mechanism acts to anyway increase the  drag. Also, despite showing drag increase, case \textit{D} features statistics that highly resemble those of the single-phase flow. Hence, only case \textit{SD} shows a more intricate two-way coupling mechanism, at play over the whole channel region.\par
\begin{figure}
   \centering
   \includegraphics[width=0.49\textwidth]{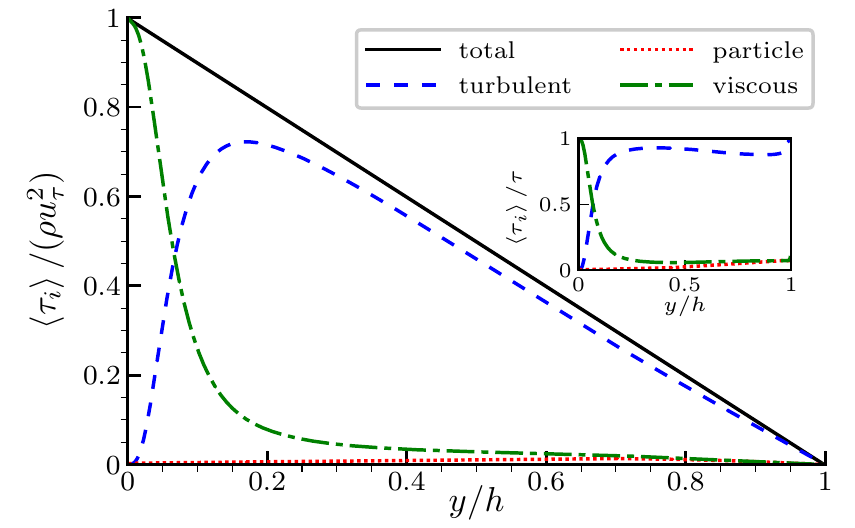}\hfill
   \includegraphics[width=0.49\textwidth]{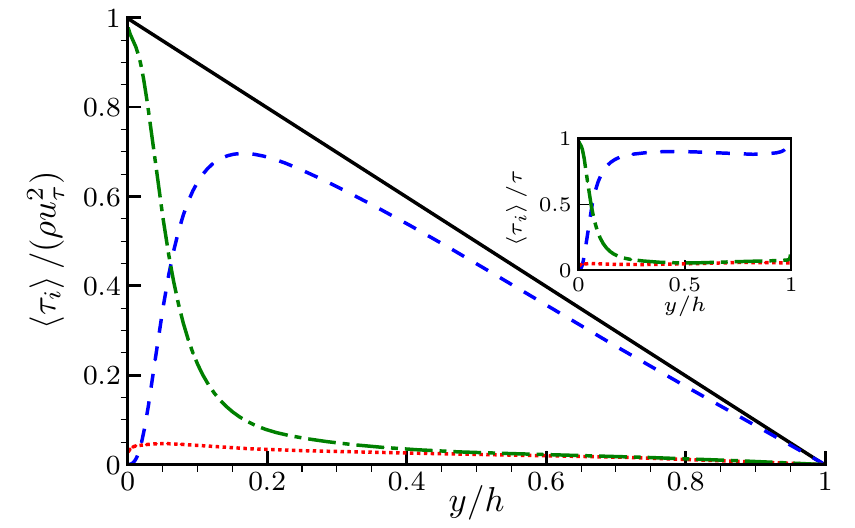}
    \put(-380, 105){\scriptsize(\textit{VD})}
    \put(-180, 105){\scriptsize~(\textit{D})}\\
   \includegraphics[width=0.49\textwidth]{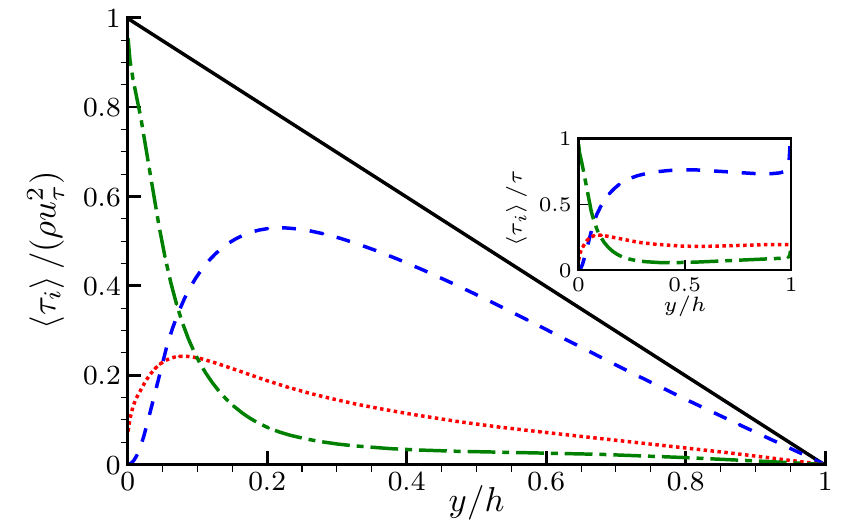}\hfill
   \includegraphics[width=0.49\textwidth]{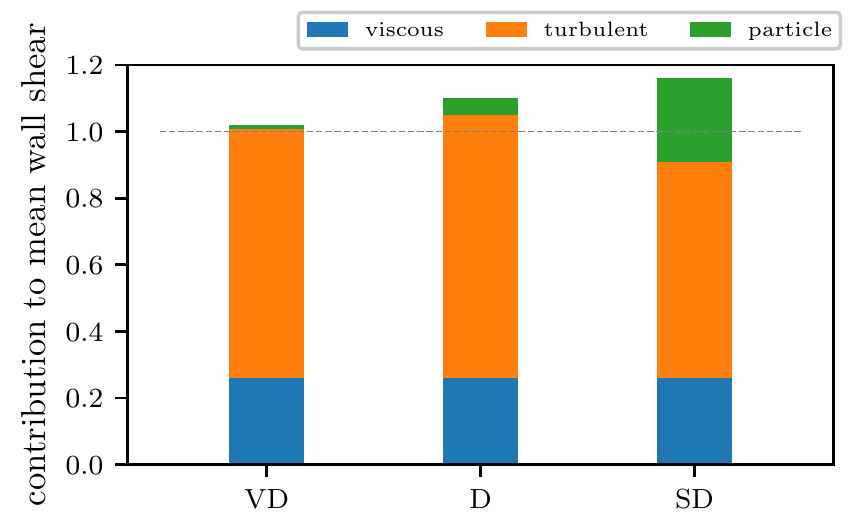}
    \put(-380, 105){\scriptsize(\textit{SD})}
    \caption{Budget of streamwise momentum as a function of the outer-scaled wall-normal distance, for the different cases. The inset shows the same budget, but normalized by the local value of the total stresses. The bottom-right panel shows the relative contribution of each term in the stress budget to the mean wall friction, normalized by that of the unladen case, i.e., $\smallint_0^1 3(1-y)\tau_i\,\mathrm{d}y/(\rho u_\tau^{2\,sph})$ \citep{Fukagata-et-al-PoF-2002,Yu-et-al-JFM-2021}, with $\tau_i$ the different contributions, and $u_\tau^{2\,sph}$ the wall friction velocity of the unladen case.}\label{fig:stress_budget}
\end{figure}

To gain  further insight, we examine the streamwise momentum balance of the fluid,
\begin{equation}
	{\tau}= {\rho}\left(1-\left\langle \phi\right\rangle \right) \left(\nu \frac{\mathrm{d}\left\langle u\right\rangle }{\mathrm{d}y} - \left\langle u_f^\prime v_f^\prime\right\rangle  \right) + \left\langle \phi\right\rangle \tau_p = \rho u_\tau^2\left(1-\frac{y}{h}\right)\mathrm{,}\label{eqn:stress_budget}
\end{equation}
where the first and second terms denote the fluid viscous and Reynolds stresses, while the last term the total particle stress $\tau_p = \tau_{p,\nu}-\rho_p\left\langle u_p^\prime v_p^\prime\right\rangle$, with $\tau_{p,\nu}$ including the viscous and collisional particle stresses. 
The data are shown in figure~\ref{fig:stress_budget}(\textit{a}-\textit{c}), where the insets depict the local relative contribution of each term to the total momentum transfer; panel (\textit{d}) reports the share of each momentum transfer mechanism to the overall drag, normalized by the overall single-phase drag. These contributions are related to the stress profiles via the FIK (Fukagata, Iwamoto \& Kasagi) identity \citep{Fukagata-et-al-PoF-2002,Yu-et-al-JFM-2021}, with the contribution of each stress mechanism $\tau_i$ given by the following weighted integral: $\smallint_0^1 3(1-y)\tau_i\,\mathrm{d}y$. Expectedly, the stress budget pertaining to case \textit{VD} resembles that of the single-phase flow, with negligible contribution from the particles. Case \textit{D} displays a noticeable, though small, contribution of the particle stresses which have a peak of about $5\%$ close to the wall and are of the order of the viscous stresses in the bulk. Finally, the total particle stresses show a significant relative contribution to the total in case \textit{SD}, of about $20\%$. Examining panel (\textit{d}) of the same figure, we note that the sum of the viscous and turbulent contributions corresponds to the single-phase drag in case \textit{VD}, with the particle stress showing an almost vanishing two-way effect. In  case \textit{D}, the sum of the viscous and turbulent stress contributions is higher than the overall single-phase drag, which is further increased by the particle stress. This indicates that the overall drag is increased by two different mechanisms: a direct one -- the particles induce an extra stress, localized near the wall where  the average slip velocity is not negligible; and an indirect effect -- the solid phase enhances the Reynolds stress and hence the turbulence. Notably, the sum of the viscous and turbulent stress contributions is smaller than the overall single-phase drag in case \textit{SD}; the particle stress, however, counterbalances this reduction and the overall drag is still higher than the corresponding unladen case. This confirms that we have turbulent drag reduction and a strong contribution of the total particle stress, which combine to a net drag increase.
\par
\begin{figure}
   \centering
   \includegraphics[width=0.49\textwidth]{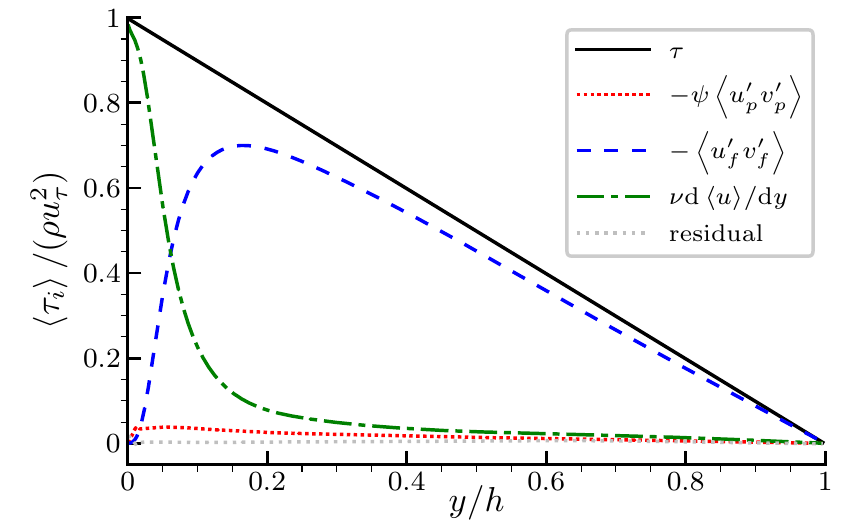}\hfill
   \includegraphics[width=0.49\textwidth]{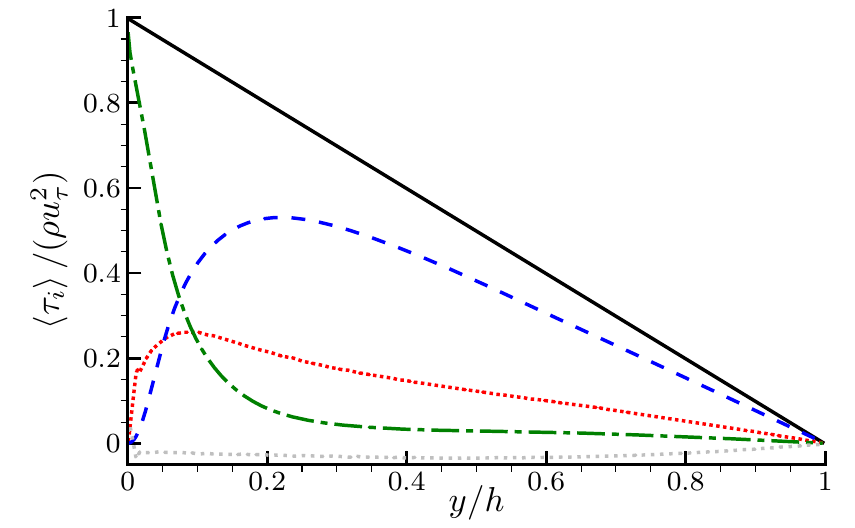}
    \put(-380, 105){\scriptsize~(\textit{D})}
    \put(-187, 105){\scriptsize(\textit{SD})}
    \caption{Budget of streamwise momentum in the two-way coupling limit, for cases \textit{D} and \textit{SD}, (see equation~\eqref{eqn:two-way}). The dashed grey line corresponds to the difference between the sum of the terms in equation~\eqref{eqn:two-way}, and the total stress, to assess the validity of the approximation. Case \textit{VD} was omitted from this figure, since given the negligible two-way coupling effect the budget is trivial, resembling that of the single-phase flow.}\label{fig:stress_budget_fp}
\end{figure}

Let us take the limit of vanishing volume fraction, $\phi \to 0$, but finite mass fraction $\psi= \rho_p\phi$, of equation eq.~\eqref{eqn:stress_budget},
\begin{equation}
\tau_{\mathrm{two-way}} = \rho\left(\nu \frac{\mathrm{d}\left\langle u\right\rangle }{\mathrm{d}y} - \left\langle u_f^\prime v_f^\prime\right\rangle  \right) - \left\langle \psi\right\rangle \left\langle u_p^\prime v_p^\prime\right\rangle\approx\rho u_\tau^2\left(1-\frac{y}{h}\right) \mathrm{.}\label{eqn:two-way}
\end{equation}
This limit corresponds to negligible particle excluded volume, but finite effects of the particle mass, i.e.\ two-way coupling conditions. In this limit, the contribution from the correlated particle fluctuations corresponds to a particle direct contribution to the drag. When plotting the terms in equations~\eqref{eqn:two-way}, see figure~\ref{fig:stress_budget_fp}, we note that the difference between the total stress and the sum of the different terms is very small, confirming that the contribution of the particle inertial shear stress to the budget is dynamically significant and increases with the mass fraction. Moreover, the ``two-way coupling'' budget explains the true nature of the turbulence attenuation for case \textit{SD}. As the mass loading is increased, the fluid Reynolds stresses progressively give in to particle Reynolds stresses, which alter the nature of the flow. A significant portion of the total stress is spent to accelerate the inertial particles, which in turn lowers the fluid Reynolds stresses. Consequently, the flow shows fluid turbulence attenuation. Unlike correlated fluid--fluid or fluid--particle velocity fluctuations, particle--particle correlated motions do not sustain near-wall turbulence, as demonstrated in \cite{Capecelatro-et-al-JFM-2018} for the fluid turbulence kinetic energy budget.\par
Since the net effect of the particle Reynolds stresses is an increased drag  for case \textit{SD}, we expect that at higher mass fractions the drag may eventually decrease -- the contribution of Reynolds stresses to the momentum transport would decrease and the flow becomes smoother and eventually laminarizes. Consequently, the correlated particle velocity fluctuations would also decrease, drastically reducing the overall flow drag. We have actually observed such a laminarizing and drag-reducing trend in a preliminary simulation of a case with the same governing parameters as \textit{SD}, but $10$ times the number of particles (note that here particle excluded volume effects may grow important). Note that, reductions of the skin friction coefficient have also been observed in the two-way coupling point-particle DNS by \cite{Vreman-JFM-2007,Zhao-et-al-PoF-2010} and in the volume-filtered Euler--Lagrange simulations of \cite{Capecelatro-et-al-JFM-2018}, with the latter study observing laminarization at sufficiently high mass fraction. In particular, \cite{Capecelatro-et-al-JFM-2018} reported that the fluid velocity fluctuations were re-energized at even higher mass fractions, due to the growing importance of turbulence-producing particle--fluid velocity correlations. New experiments and fully resolved simulations are hence needed to  further support and explain these  observations at higher mass fractions.
\section{Conclusions}\label{sec:conclusions}
We have used particle-resolved DNS to study the near-wall turbulence modulation by small inertial particles. Three cases have been considered with volume fractions progressively increased by one order of magnitude, chosen in the one-way and two-way coupling regimes. The two densest cases show a non-negligible turbulence modulation, however, of a fundamentally different nature. The case with $\Psi=3.4\%$ ($\Phi=0.034\%$) features drag increase due to increased inertia near the wall, with increased Reynolds stresses due to particles travelling at high particle-to-fluid slip velocity, and dynamics close to that of the single-phase flow at slightly higher Reynolds number. Increasing the volume and mass fraction by a factor of $10$,  the particle presence is felt over the entire channel, however, the relative drag increase is significantly reduced with respect to that of the previous configuration. The streamwise momentum balance in the two-way coupling limit of vanishing volume fraction, but finite mass fraction, reveals  that correlated particle fluctuations seize a larger share of the total stresses as the mass fraction increases. This results in a reduction of the fluid Reynolds stresses, and consequently of the turbulent drag. For the parameter setting considered here, this turbulent drag reduction is counterbalanced by the contribution of correlated particle velocity fluctuations; however, a further increase in mass fraction may lead to a net drag reduction, something which has been noticed in previous particle-modelled DNS, and should inspire future experimental and particle-resolved numerical studies.\\

This work was supported by the European Research Council grant no.\ ERC-2013-CoG-616186, TRITOS, the grant BIRD192032/19 from the University of Padova, and the University of Iceland Recruitment Fund grant no.~1515-151341, \emph{TURBBLY}. We acknowledge the computing time provided by SNIC (Swedish National Infrastructure for Computing), and PRACE for awarding us access to the supercomputer Marconi, based in Italy at {CINECA} under project 2017174185 -- \textit{DILPART}.\\

Declaration of Interests. The authors report no conflict of interest.

\bibliographystyle{jfm}

\bibliography{bibfile.bib}

\begin{thebibliography}{33}
\expandafter\ifx\csname natexlab\endcsname\relax\def\natexlab#1{#1}\fi
\def\au#1{#1} \def\ed#1{#1} \def\yr#1{#1}\def\at#1{#1}\def\jt#1{\textit{#1}}
  \def\bt#1{#1}\def\bvol#1{\textbf{#1}} \def\vol#1{#1} \def\pg#1{#1}
  \def\publ#1{#1}\def\arxiv#1{#1}\def\org#1{#1}\def\st#1{\textit{#1}}

\bibitem[Balachandar \& Eaton(2010)]{Balachandar-and-Eaton-ARFM-2010}
{\sc \au{Balachandar, S} \& \au{Eaton, John~K}} \yr{2010}  \at{Turbulent
  dispersed multiphase flow}.  \jt{Annu.\ Rev.\ Fluid Mech.}  \bvol{42},
  \pg{111--133}.

\bibitem[Battista {\em et~al.\/}(2019)Battista, Mollicone, Gualtieri, Messina
  \& Casciola]{Battista-et-al-JFM-2019}
{\sc \au{Battista, F}, \au{Mollicone, J-P}, \au{Gualtieri, P}, \au{Messina, R}
  \& \au{Casciola, CM}} \yr{2019}  \at{Exact regularised point particle (erpp)
  method for particle-laden wall-bounded flows in the two-way coupling regime}.
   \jt{J. Fluid Mech.}  \bvol{878},  \pg{420--444}.

\bibitem[Breugem(2012)]{Breugem-JCP-2012}
{\sc \au{Breugem, W-P}} \yr{2012}  \at{A second-order accurate immersed
  boundary method for fully resolved simulations of particle-laden flows}.
  \jt{J. Comput. Phys.}  \bvol{231}~(13),  \pg{4469--4498}.

\bibitem[Capecelatro {\em et~al.\/}(2018)Capecelatro, Desjardins \&
  Fox]{Capecelatro-et-al-JFM-2018}
{\sc \au{Capecelatro, Jesse}, \au{Desjardins, Olivier} \& \au{Fox, Rodney~O.}}
  \yr{2018}  \at{On the transition between turbulence regimes in particle-laden
  channel flows}.  \jt{J. Fluid Mech.}  \bvol{845},  \pg{499--519}.

\bibitem[Costa(2018)]{Costa-CAMWA-2018}
{\sc \au{Costa, P}} \yr{2018}  \at{A fft-based finite-difference solver for
  massively-parallel direct numerical simulations of turbulent flows}.
  \jt{Comput. Math. Appl.}  \bvol{76}~(8),  \pg{1853 -- 1862}.

\bibitem[Costa {\em et~al.\/}(2015)Costa, Boersma, Westerweel \&
  Breugem]{Costa-et-al-PRE-2015}
{\sc \au{Costa, P}, \au{Boersma, BJ}, \au{Westerweel, J} \& \au{Breugem, W-P}}
  \yr{2015}  \at{Collision model for fully resolved simulations of flows laden
  with finite-size particles}.  \jt{Phys. Rev. E}  \bvol{92}~(5),  \pg{053012}.

\bibitem[Costa {\em et~al.\/}(2020)Costa, Brandt \&
  Picano]{Costa-et-al-JFM-2020}
{\sc \au{Costa, P}, \au{Brandt, L} \& \au{Picano, F}} \yr{2020}
  \at{Interface-resolved simulations of small inertial particles in turbulent
  channel flow}.  \jt{J. Fluid Mech.}  \bvol{883}.

\bibitem[Crowe {\em et~al.\/}(1977)Crowe, Sharma \&
  Stock]{Crowe-et-al-JFE-1977}
{\sc \au{Crowe, CT}, \au{Sharma, MPt} \& \au{Stock, DE}} \yr{1977}  \at{The
  particle-source-in cell (psi-cell) model for gas-droplet flows}.  \jt{J.\
  Fluids Engng.}  \bvol{99}~(2),  \pg{325--332}.

\bibitem[Fornari {\em et~al.\/}(2016)Fornari, Formenti, Picano \&
  Brandt]{Fornari-et-al-PoF-2016}
{\sc \au{Fornari, W}, \au{Formenti, A}, \au{Picano, F} \& \au{Brandt, L}}
  \yr{2016}  \at{The effect of particle density in turbulent channel flow laden
  with finite size particles in semi-dilute conditions}.  \jt{Phys. Fluids}
  \bvol{28}~(3),  \pg{033301}.

\bibitem[Fr{\"o}hlich {\em et~al.\/}(2018)Fr{\"o}hlich, Schneiders, Meinke \&
  Schr{\"o}der]{Frohlich-et-al-FTC-2018}
{\sc \au{Fr{\"o}hlich, K}, \au{Schneiders, L}, \au{Meinke, M} \&
  \au{Schr{\"o}der, W}} \yr{2018}  \at{Validation of lagrangian two-way coupled
  point-particle models in large-eddy simulations}.  \jt{Flow Turbul.\
  Combust.}  \bvol{101}~(2),  \pg{317--341}.

\bibitem[Fukagata {\em et~al.\/}(2002)Fukagata, Iwamoto \&
  Kasagi]{Fukagata-et-al-PoF-2002}
{\sc \au{Fukagata, Koji}, \au{Iwamoto, Kaoru} \& \au{Kasagi, Nobuhide}}
  \yr{2002}  \at{Contribution of reynolds stress distribution to the skin
  friction in wall-bounded flows}.  \jt{Phys. Fluids}  \bvol{14}~(11),
  \pg{L73--L76}.

\bibitem[Gualtieri {\em et~al.\/}(2015)Gualtieri, Picano, Sardina \&
  Casciola]{Gualtieri-et-al-JFM-2015}
{\sc \au{Gualtieri, P}, \au{Picano, F}, \au{Sardina, G} \& \au{Casciola, CM}}
  \yr{2015}  \at{Exact regularized point particle method for multiphase flows
  in the two-way coupling regime}.  \jt{J.\ Fluid Mech.}  \bvol{773},
  \pg{520--561}.

\bibitem[Horne \& Mahesh(2019)]{Horne-and-Mahesh-JCP-2019}
{\sc \au{Horne, WJ} \& \au{Mahesh, K}} \yr{2019}  \at{A massively-parallel,
  unstructured overset method to simulate moving bodies in turbulent flows}.
  \jt{J. Comput. Phys.}  \bvol{397},  \pg{108790}.

\bibitem[Horwitz \& Mani(2016)]{Horwitz-and-Mani-JCP-2016}
{\sc \au{Horwitz, JAK} \& \au{Mani, A}} \yr{2016}  \at{Accurate calculation of
  stokes drag for point--particle tracking in two-way coupled flows}.  \jt{J.\
  Comput.\ Phys.}  \bvol{318},  \pg{85--109}.

\bibitem[Ireland \& Desjardins(2017)]{Ireland-and-Desjardins-JCP-2017}
{\sc \au{Ireland, PJ} \& \au{Desjardins, O}} \yr{2017}  \at{Improving particle
  drag predictions in euler--lagrange simulations with two-way coupling}.
  \jt{J.\ Comput.\ Phys.}  \bvol{338},  \pg{405--430}.

\bibitem[Kim \& Moin(1985)]{Kim-and-Moin-JCP-1985}
{\sc \au{Kim, J} \& \au{Moin, P}} \yr{1985}  \at{Application of a
  fractional-step method to incompressible navier-stokes equations}.  \jt{J.
  Comput. Phys.}  \bvol{59}~(2),  \pg{308--323}.

\bibitem[Kuerten \& Vreman(2016)]{Kuerten-and-Vreman-IJMF-2016}
{\sc \au{Kuerten, J.G.M.} \& \au{Vreman, A.W.}} \yr{2016}  \at{Collision
  frequency and radial distribution function in particle-laden turbulent
  channel flow}.  \jt{Int. J. Multiphase Flow}  \bvol{87},  \pg{66--79}.

\bibitem[Kulick {\em et~al.\/}(1994)Kulick, Fessler \&
  Eaton]{Kulick-et-al-JFM-1994}
{\sc \au{Kulick, Jonathan~D}, \au{Fessler, John~R} \& \au{Eaton, John~K}}
  \yr{1994}  \at{Particle response and turbulence modification in fully
  developed channel flow}.  \jt{J. Fluid Mech.}  \bvol{277},  \pg{109--134}.

\bibitem[Marchioli \& Soldati(2002)]{Marchioli-and-Soldati-JFM-2003}
{\sc \au{Marchioli, Cristian} \& \au{Soldati, Alfredo}} \yr{2002}
  \at{Mechanisms for particle transfer and segregation in a turbulent boundary
  layer}.  \jt{J. Fluid Mech.}  \bvol{468},  \pg{283--315}.

\bibitem[Mehrabadi {\em et~al.\/}(2018)Mehrabadi, Horwitz, Subramaniam \&
  Mani]{Mehrabadi-et-al-JFM-2018}
{\sc \au{Mehrabadi, M}, \au{Horwitz, JAK}, \au{Subramaniam, S} \& \au{Mani, A}}
  \yr{2018}  \at{A direct comparison of particle-resolved and point-particle
  methods in decaying turbulence}.  \jt{J.\ Fluid Mech.}  \bvol{850},
  \pg{336--369}.

\bibitem[Pakseresht {\em et~al.\/}(2020)Pakseresht, Esmaily \&
  Apte]{Pakseresht-and-Apte-JCP-2020}
{\sc \au{Pakseresht, Pedram}, \au{Esmaily, Mahdi} \& \au{Apte, Sourabh~V.}}
  \yr{2020}  \at{A correction scheme for wall-bounded two-way coupled
  point-particle simulations}.  \jt{J. Comput. Phys.}  \bvol{420},
  \pg{109711}.

\bibitem[Picano {\em et~al.\/}(2015)Picano, Breugem \&
  Brandt]{Picano-et-al-JFM-2015}
{\sc \au{Picano, F}, \au{Breugem, W-P} \& \au{Brandt, L}} \yr{2015}
  \at{Turbulent channel flow of dense suspensions of neutrally buoyant
  spheres}.  \jt{J.\ Fluid Mech.}  \bvol{764},  \pg{463--487}.

\bibitem[Richter(2015)]{Richter-PoF-2015}
{\sc \au{Richter, David~H}} \yr{2015}  \at{Turbulence modification by inertial
  particles and its influence on the spectral energy budget in planar couette
  flow}.  \jt{Phys. Fluids}  \bvol{27}~(6),  \pg{063304}.

\bibitem[Rouson \& Eaton(2001)]{Rouson-and-Eaton-JFM-2001}
{\sc \au{Rouson, DWI} \& \au{Eaton, JK}} \yr{2001}  \at{On the preferential
  concentration of solid particles in turbulent channel flow}.  \jt{J. of Fluid
  Mech.}  \bvol{428},  \pg{149}.

\bibitem[Sardina {\em et~al.\/}(2012)Sardina, Schlatter, Brandt, Picano \&
  Casciola]{Sardina-et-al-JFM-2012}
{\sc \au{Sardina, G}, \au{Schlatter, P}, \au{Brandt, L}, \au{Picano, F} \&
  \au{Casciola, CM}} \yr{2012}  \at{Wall accumulation and spatial localization
  in particle-laden wall flows}.  \jt{J. Fluid Mech.}  \bvol{699}~(1),
  \pg{50--78}.

\bibitem[Schneiders {\em et~al.\/}(2017)Schneiders, Meinke \&
  Schr{\"o}der]{Schneiders-et-al-JFM-2017}
{\sc \au{Schneiders, L}, \au{Meinke, M} \& \au{Schr{\"o}der, W}} \yr{2017}
  \at{Direct particle--fluid simulation of kolmogorov-length-scale size
  particles in decaying isotropic turbulence}.  \jt{J.\ Fluid Mech.}
  \bvol{819},  \pg{188--227}.

\bibitem[Sundaram \& Collins(1997)]{Sundaram-and-Collins-JFM-1997}
{\sc \au{Sundaram, S} \& \au{Collins, LR}} \yr{1997}  \at{Collision statistics
  in an isotropic particle-laden turbulent suspension. part 1. direct numerical
  simulations}.  \jt{J. Fluid Mech.}  \bvol{335},  \pg{75--109}.

\bibitem[Tiederman {\em et~al.\/}(1985)Tiederman, Luchik \&
  Bogard]{Tiederman-et-al-JFM-1985}
{\sc \au{Tiederman, WG}, \au{Luchik, Thomas~S} \& \au{Bogard, DG}} \yr{1985}
  \at{Wall-layer structure and drag reduction}.  \jt{J. Fluid Mech.}
  \bvol{156},  \pg{419--437}.

\bibitem[Uhlmann(2005)]{Uhlmann-JCP-2005}
{\sc \au{Uhlmann, M}} \yr{2005}  \at{An immersed boundary method with direct
  forcing for the simulation of particulate flows}.  \jt{J. Comput. Phys.}
  \bvol{209}~(2),  \pg{448--476}.

\bibitem[Vreman(2007)]{Vreman-JFM-2007}
{\sc \au{Vreman, A.~W.}} \yr{2007}  \at{Turbulence characteristics of
  particle-laden pipe flow}.  \jt{J. Fluid Mech.}  \bvol{584},  \pg{235--279}.

\bibitem[Wang {\em et~al.\/}(2019)Wang, Fong, Coletti, Capecelatro \&
  Richter]{Wang-et-al-IJMF-2019}
{\sc \au{Wang, G}, \au{Fong, KO}, \au{Coletti, F}, \au{Capecelatro, J} \&
  \au{Richter, DH}} \yr{2019}  \at{Inertial particle velocity and distribution
  in vertical turbulent channel flow: A numerical and experimental comparison}.
   \jt{Int.\ J.\ Multiphase Flow}  \bvol{120},  \pg{103105}.

\bibitem[Yu {\em et~al.\/}(2021)Yu, Xia, Guo \& Lin]{Yu-et-al-JFM-2021}
{\sc \au{Yu, Zhaosheng}, \au{Xia, Yan}, \au{Guo, Yu} \& \au{Lin, Jianzhong}}
  \yr{2021}  \at{Modulation of turbulence intensity by heavy finite-size
  particles in upward channel flow}.  \jt{J. Fluid Mecha.}  \bvol{913}.

\bibitem[Zhao {\em et~al.\/}(2010)Zhao, Andersson \&
  Gillissen]{Zhao-et-al-PoF-2010}
{\sc \au{Zhao, LH}, \au{Andersson, H~I} \& \au{Gillissen, JJJ}} \yr{2010}
  \at{Turbulence modulation and drag reduction by spherical particles}.
  \jt{Phys. Fluids}  \bvol{22}~(8),  \pg{081702}.

\end{thebibliography}

\end{document}